\documentclass[%
reprint,
superscriptaddress,
 amsmath,amssymb,
 aps,
 prl,
]{revtex4-2}
\usepackage{float}
\usepackage{graphicx}
\usepackage{dcolumn}
\usepackage{bm}
\usepackage{hyperref}
\usepackage{lineno}
\usepackage{xcolor}

\begin{document}


\title{Nuclear Geometry-Driven Femtoscopic Correlations in Light-Ion Collisions}

\author{D\'aniel Kincses}
\email{kincses@ttk.elte.hu}
\affiliation{%
 ELTE E\"otv\"os Lor\'and University, P\'azm\'any P\'eter s\'et\'any 1/A, H-1117 Budapest, Hungary
}%
\affiliation{%
 Faculty of Nuclear Sciences and Physical Engineering, Czech Technical University in Prague, Břehová 7, 11519, Prague 1, Czech Republic
}%
\author{Bithika Karmakar}
\author{Iurii Karpenko}
\affiliation{%
 Faculty of Nuclear Sciences and Physical Engineering, Czech Technical University in Prague, Břehová 7, 11519, Prague 1, Czech Republic
}%
\author{Boris Tom\'a\v{s}ik}
\affiliation{%
 Faculty of Nuclear Sciences and Physical Engineering, Czech Technical University in Prague, Břehová 7, 11519, Prague 1, Czech Republic
}%
\affiliation{
Fakulta pr\'irodn\'ych vied, Univerzita Mateja Bela, Tajovsk\'eho 40, 97401 Bansk\'a Bystrica, Slovakia
}%
\begin{abstract}

Recent anisotropic flow measurements in $^{16}\rm O{+}^{16}\rm O$ and $^{20}\rm Ne{+}^{20}\rm Ne$ collisions at a center-of-mass energy per nucleon pair of 5.36~TeV provide evidence for nuclear geometry-driven collectivity. In this Letter, we investigate the femtoscopic signatures of nuclear geometry in these light-ion collision systems, using a hybrid simulation framework that couples \textit{ab initio} nuclear configurations to viscous hydrodynamics and hadronic transport. We show that the $\rm Ne{+}Ne/O{+}O$ ratio of the freeze-out eccentricity, extracted from azimuthally sensitive femtoscopy, reaches 1.5--1.8 in central collisions for deformed nuclear configurations, compared with about unity for a spherical baseline. This enhancement is significantly larger than that of elliptic flow in the same simulations, and remains stable when pre-equilibrium free streaming is omitted, even though the absolute source radii change considerably. Confronting these predictions with future data will provide new, complementary constraints on nuclear structure and initial-condition modeling.

\end{abstract}
\vspace{-10pt}

\keywords{Nuclear structure, alpha-clustering, Lévy-stable distribution, high-energy physics, heavy-ion physics, femtoscopy}
\maketitle

\paragraph{\label{sec:intro}Introduction.}Recently, a new avenue has emerged in nuclear structure studies via relativistic collisions of light nuclei, forging yet another intersection between low- and high-energy nuclear physics~\cite{Mazeliauskas:2026pqy}. The recent Large Hadron Collider (LHC) data taking featuring $^{16}\rm O{+}^{16}\rm O$ and $^{20}\rm Ne{+}^{20}\rm Ne$ collisions at a center-of-mass energy per nucleon pair of $\sqrt{s_{NN}} = 5.36$~TeV provides a unique opportunity to probe intrinsic nuclear deformation and $\alpha$-clustering, such as the predicted tetrahedral structure of $^{16}\rm O$ and the bowling-pin geometry of $^{20}\rm Ne$~\cite{Giacalone:2024luz}. Indeed, recent anisotropic flow ($v_n$) measurements by LHC experiments have observed such signatures~\cite{ALICE:2025luc,CMS:2025tga,ATLAS:2025nnt}, providing evidence for nuclear geometry-driven collectivity.

Anisotropic flow reflects the hydrodynamic conversion of initial spatial asymmetries into final-state momentum anisotropies~\cite{Niemi:2012aj}. An imprint of the initial geometry is, however, also encoded in the spatial configuration of the particle-emitting source at the moment when the medium is converted into hadrons. This imprint is carried through the hadronic rescattering stage to kinetic freeze-out, when the hadrons cease to interact. Two-particle femtoscopy (HBT interferometry) provides experimental access to the space-time extent of this emission source at kinetic freeze-out~\cite{Lisa:2005dd}. In particular, the azimuthal dependence of the femtoscopic radii relative to the event plane gives access to the freeze-out eccentricity $\varepsilon_{\rm side}$~\cite{PhysRevC.66.044903,Retiere:2003kf,STAR:2014shf,ALICE:2017gxt}. Since the radii measure homogeneity regions shaped by both the geometry and the collective flow, $\varepsilon_{\rm side}$ is not a direct image of the initial state. Analytical parametrisations of fluid dynamical evolution and freeze-out process~\cite{Csanad:2008af,Cimerman:2017lmm,Lokos:2016fze}, 
however, show that azimuthally sensitive femtoscopy and elliptic flow depend on the initial spatial anisotropy and on the flow field in different combinations; measuring both observables can therefore help disentangle geometric and dynamical effects. Furthermore, global Bayesian analyses of heavy-ion collisions have demonstrated that femtoscopic radii have significant constraining power on the equation of state and early-time dynamics~\cite{Pratt:2015zsa,Sangaline:2015isa}. As such inference techniques are increasingly applied to extract nuclear deformation parameters from high-energy data~\cite{Giacalone:2026fat}, future global analyses of light-ion collisions could strongly benefit from including femtoscopic observables.

Femtoscopy has only begun to be explored as a probe of nuclear structure, with studies so far based on a multi-phase transport (AMPT) model and mostly relying on Gaussian-based source parametrizations~\cite{He:2020jzd,Xi:2026vrp}. In contrast, the pion pair source in high-energy collisions exhibits a power-law tail, well described by L\'evy-stable distributions~\cite{Kincses:2024lnv,PHENIX:2017ino,PHENIX:2024vjp,CMS:2023xyd,NA61SHINE:2023qzr}. In Ref.~\cite{Kincses:2025iaf}, a three-dimensional L\'evy-stable shape analysis was used in an AMPT simulation to show that $\varepsilon_{\rm side}$ is sensitive to the deformed shape of $^{20}\rm Ne$ in $^{208}\rm Pb{+}^{20}\rm Ne$ and $^{208}\rm Pb{+}^{16}\rm O$ collisions at ${\sqrt{s_{NN}} = 68.5}$~GeV, relevant for the LHCb fixed-target program. Extending this approach to $^{16}\rm O{+}^{16}\rm O$ and $^{20}\rm Ne{+}^{20}\rm Ne$ collisions at ${\sqrt{s_{NN}} = 5.36}$~TeV requires a realistic treatment of both the fluid-dynamical expansion and the hadronic phase.

In this Letter, we employ a hybrid simulation framework for this purpose. We couple \textit{ab initio} nucleon configurations~\cite{Giacalone:2024luz} to the Reduced Thickness Event-by-event Nuclear Topology (T$_{\text{R}}$ENTo~\cite{Moreland:2014oya}) initial condition model, followed by viscous hydrodynamics (VISHNew~\cite{Song:2007ux,Song:2008si}) modeling the collective expansion of the quark-gluon plasma, and by the UrQMD hadronic transport afterburner~\cite{Bleicher:1999xi,Bass:1998ca}, which describes the rescattering and resonance decays governing the freeze-out process. Using this framework, we predict the azimuthally sensitive three-dimensional femtoscopic radii in $^{16}\rm O{+}^{16}\rm O$ and $^{20}\rm Ne{+}^{20}\rm Ne$ collisions at $\sqrt{s_{NN}} = 5.36$~TeV. We find that the $\rm Ne{+}Ne/O{+}O$ ratio of the freeze-out eccentricity is strongly enhanced in central collisions for deformed nuclear configurations, significantly more than the corresponding ratio of elliptic flow. Together with future measurements, these observables will provide additional constraints on nuclear structure calculations and initial-condition modeling.

\paragraph{\label{sec:met}Methods.}To study the effects of nuclear geometry in light-ion collisions, we use randomly oriented nucleon configurations from the nuclear lattice effective field theory (NLEFT) and projected generator coordinate method (PGCM) \textit{ab initio} calculations~\cite{Giacalone:2024luz}, and a spherical Woods-Saxon (W-S) baseline~\cite{Giacalone:2024ixe}. 
The T$_{\text{R}}$ENTo initial conditions are followed by pre-equilibrium free streaming~\cite{Liu:2015nwa,Broniowski:2008qk} until $\tau=0.48$~fm/$c$~\cite{Moreland:2018gsh} and by boost-invariant viscous hydrodynamical evolution. To test the sensitivity to the pre-equilibrium dynamics, we also generated a PGCM ensemble without free streaming (PGCM w/o FS), with hydrodynamics starting also at ${\tau=0.48}$~fm/$c$. The T$_{\text{R}}$ENTo normalization was tuned separately for each configuration to reproduce the measured charged-particle multiplicities at midrapidity~\cite{Modak:2026ymk}. For other T$_{\text{R}}$ENTo input parameters, we use the maximum a posteriori (MAP) values found in  Ref.~\cite{Moreland:2018gsh} with nucleon width 0.8~fm and subnucleon width 0.4~fm.
For each configuration, we simulated $10^5$ minimum-bias hydrodynamic evolutions for both $^{20}\rm Ne{+}^{20}\rm Ne$ and $^{16}\rm O{+}^{16}\rm O$ collisions. Each freeze-out hypersurface was sampled 50 times via the Cooper-Frye formula~\cite{Cimerman:2021gwf}, and each sample was passed to UrQMD~\cite{Bleicher:1999xi,Bass:1998ca}. Events were sorted into 5\% centrality classes by the midrapidity charged-particle multiplicity averaged over the 50 samples of each hydrodynamic event.

The second-order event plane $\Psi_2$ is determined in each oversampled event from charged pions, kaons, and protons in the pseudorapidity ranges of $-2.5<\eta<-0.5$ and $0.5<\eta<2.5$, following Eqs.~(3)--(5) of Ref.~\cite{Kincses:2025iaf}. Using the space-time coordinates and momenta of particles at their last interaction~\cite{Kincses:2024lnv,Kincses:2025iaf}, we select charged pions with $|\eta|<0.5$ and $0.15<p_T<1.5$~GeV/$c$, and form same-event pairs in bins of average transverse momentum $k_T$ and of $\varphi_{\rm pair}-\Psi_2$, with $\Psi_2$ taken from the hemisphere opposite to the pair rapidity. As in Refs.~\cite{Kincses:2024lnv,Kincses:2025izu,Kincses:2025iaf}, we keep pairs with $Q_{\rm LCMS}<\sqrt{(0.25~{\rm GeV})\,m_T}$, relevant for experimental correlation measurements, where $Q_{\rm LCMS}$ is the relative momentum in the longitudinally comoving system (LCMS)~\cite{PHENIX:2017ino} and $m_T=\sqrt{m_\pi^2+k_T^2}$ is the pair average transverse mass. For these pairs, we compute the pair-distance distribution $D(\vec\rho)$ in the LCMS~\cite{Kincses:2024lnv}, using out-side-long coordinates~\cite{Grassberger:1976au,Podgoretsky:1982xu,Bertsch:1993nx,Pratt:1995wm}, where “out” is along the direction of $k_T$, “long” is the beam direction, and “side” is perpendicular to the rest. 

The core equation of femtoscopy describes the connection between the relative momentum correlation and the pair source~\cite{Goldhaber:1960sf,Kopylov:1974th,Podgoretsky:1989bp,Koonin:1977fh,Yano:1978gk,Lednicky:1981su,Pratt:1984su,Pratt:1997pw}:
\begin{align}
C_2(\vec{q},\vec{K}) &= \int d^3\vec{\rho}\, D_{\vec{K}}(\vec{\rho})\, |\psi_{\vec{q}}(\vec{\rho})|^2 \nonumber \\
&= 1 + \int d^3\vec{\rho}\, D_{\vec{K}}(\vec{\rho}) \cos(\vec{q}\cdot\vec{\rho}),
\label{e:C2D}
\end{align}
where $\vec q$ and $\vec K$ are the relative and average pair momenta, $D$ is integrated over the time separation in the pair rest frame, and the second equality holds for the symmetrized pair wave function of identical bosons without final-state interactions~\cite{Nagy:2023zbg,Nagy:2026mvg}. Previous studies~\cite{Kincses:2024lnv,Kincses:2025izu,Kincses:2025iaf} found, in agreement with experiments~\cite{PHENIX:2017ino,PHENIX:2024vjp,CMS:2023xyd,NA61SHINE:2023qzr}, that the pion pair source is well described by an elliptically contoured three-dimensional L\'evy-stable distribution. Since L\'evy-stable distributions must be evaluated numerically (which is computationally demanding in three dimensions), these studies fitted one-dimensional projections of $D_{\vec K}(\vec\rho)$. Here, we instead compute $C_2(\vec q)$ via Eq.~\eqref{e:C2D} by a fast Fourier transform (FFT) of $D_{\vec K}(\vec\rho)$ constructed on a three-dimensional grid. In Fourier space, the L\'evy-stable form is analytic, which enables a fully three-dimensional fit; moreover, fitting $C_2(\vec q)$ weights the different regions of the source in the same way as experimental correlation analyses do:
\begin{align}
C_2(\vec{q}) = 1 + \lambda \exp\left(-\left(\vec{q}^T \mathbf{R}^2 \vec{q}\right)^{\alpha/2}\right),
\label{e:Cq3D}
\end{align}
where $\lambda$ is the correlation strength, $\alpha$ is the L\'evy exponent, and $\mathbf{R}^2$ is the matrix of squared L\'evy scale parameters. In the Bertsch--Pratt frame, $\vec{q}^{\,T}\mathbf{R}^2\vec{q} = q_{\rm out}^2 R_{\rm out}^2 + q_{\rm side}^2 R_{\rm side}^2 + q_{\rm long}^2 R_{\rm long}^2 + 2 q_{\rm out} q_{\rm side} R_{\rm outside}^2$; the out-long and side-long terms vanish for symmetric systems at midrapidity~\cite{PhysRevC.66.044903}. The out-side term reflects the orientation of the emission region relative to the pair momentum, with additional contributions from collective flow~\cite{Csanad:2008af,Cimerman:2017lmm,Lokos:2016fze}. Fitting each azimuthal bin independently yields the $\varphi_{\rm pair}-\Psi_2$ dependence of the radii, which we parametrize as
\begin{align}
R_{i}^2(\varphi_{\rm pair} {-} \Psi_2) &= R_{i,0}^2 + 2 R_{i,2}^2 \cos(2(\varphi_{\rm pair} {-} \Psi_2)), \nonumber \\
R_{\rm outside}^2(\varphi_{\rm pair} {-} \Psi_2) &= 2 R_{{\rm outside},2}^2 \sin(2(\varphi_{\rm pair} {-} \Psi_2)),
\label{e:oscillation}
\end{align}
where $i \in \{{\rm out}, {\rm side}, {\rm long}\}$. We define the eccentricities $\varepsilon_i = 2R_{i,2}^2 / R_{i,0}^2$. Here $\varepsilon_{\rm side}$ corresponds to the freeze-out eccentricity of Refs.~\cite{Kincses:2025iaf,STAR:2014shf,ALICE:2017gxt} and is most directly related to the transverse shape of the emitting region~\cite{Retiere:2003kf}, while $\varepsilon_{\rm out}$ also contains contributions from the emission duration and flow. Since nuclear deformation primarily affects the transverse radii, we focus on $\varepsilon_{\rm out}$, $\varepsilon_{\rm side}$, and the cross-term amplitude $2R_{{\rm outside},2}^2$, and evaluate their $\rm Ne{+}Ne/O{+}O$ ratios to reduce the sensitivity to the common bulk evolution. The uncertainties of the ratios are obtained by propagating the statistical uncertainties of the individual quantities. Details of the fitting procedure and of the uncertainty estimation are given in the Supplemental Material.

Our simulated correlation functions contain only the quantum-statistical (Bose--Einstein) effect, while final-state Coulomb and strong interactions are neglected. This allows the use of the analytic parametrization of Eq.~\eqref{e:Cq3D}, which keeps the large number of multi-differential fits numerically stable. 
Experimental analyses remove the Coulomb repulsion when extracting the source parameters, so our results correspond to the source targeted by such measurements. While Coulomb corrections for three-dimensional sources were historically challenging, they can now be calculated self-consistently not just for Gaussian but even for L\'evy-stable sources~\cite{Nagy:2026mvg}, opening the way to precision azimuthally sensitive analyses of experimental data. 

\paragraph{\label{sec:res}Results and discussion.}

\begin{figure}
    \centering
    \includegraphics[width=\linewidth]{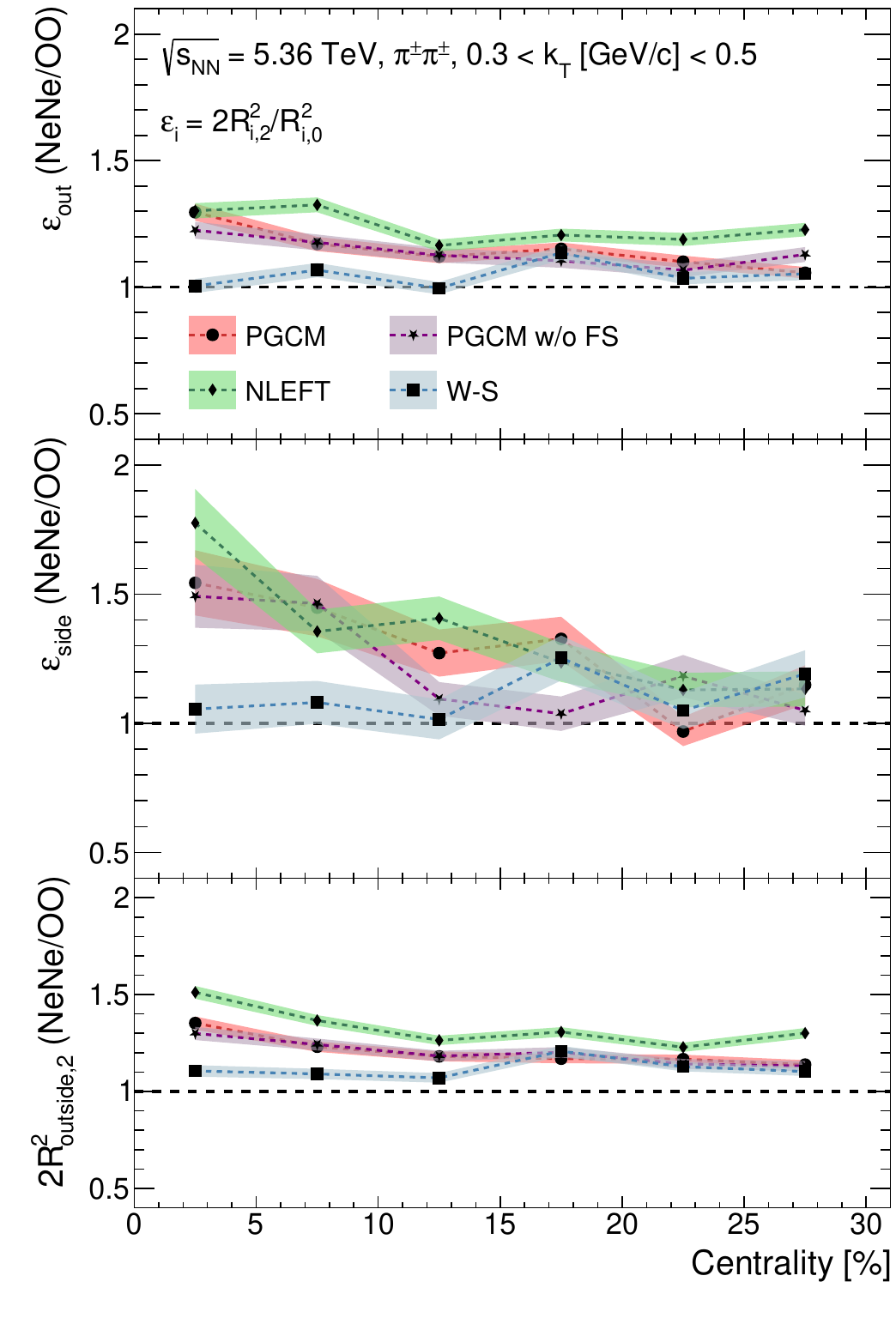}
    \caption{Centrality dependent ratios of the second-order Fourier terms of the azimuthally sensitive femtoscopic radii in Ne+Ne to O+O collisions. The panels display the relative eccentricities $\varepsilon_{\text{out}}$ (top) and $\varepsilon_{\text{side}}$ (middle), alongside the cross-term $2R_{\text{outside},2}^2$ (bottom). On each panel, the initial nucleon configuration setups are shown with different marker shapes and colored bands representing  uncertainties propagated from those of the individual quantities.}
    \label{fig:centdepratios}
\end{figure}

Fig.~\ref{fig:centdepratios} presents the centrality dependence of the $\rm Ne{+}Ne/O{+}O$ ratios of the freeze-out eccentricities $\varepsilon_{\rm out}$ and $\varepsilon_{\rm side}$, and of the cross-term amplitude $2R_{{\rm outside},2}^{2}$, for $0.3 < k_T < 0.5$~GeV/$c$, the average transverse momentum range containing most of the pair statistics. The spherical W-S baseline yields ratios close to unity for $\varepsilon_{\rm out}$ and $\varepsilon_{\rm side}$ over the whole centrality range, and about 1.1 for the unnormalized $2R_{{\rm outside},2}^{2}$. Without intrinsic deformation, the freeze-out eccentricities of the two systems are thus similar, indicating that generic differences between the $A=20$ and $A=16$ systems in size and expansion largely cancel in the ratio. A significant deviation from unity therefore signals a difference in intrinsic nuclear geometry that survives to freeze-out.

The \textit{ab initio} configurations instead show a clear enhancement in central collisions, largest in the 0--5\% class, where the eccentricity is not dominated by the impact-parameter-driven overlap geometry. Here, the bowling-pin shape of $^{20}$Ne leads to a more elongated transverse freeze-out region than the tetrahedral shape of $^{16}$O, for which $\varepsilon_{\rm side}$ is nearly independent of the nuclear configuration (see Supplemental Material). The $\varepsilon_{\rm side}$ ratio reaches about 1.5 for PGCM and 1.8 for NLEFT, while the $\varepsilon_{\rm out}$ ratio is about 1.3 and the $2R_{{\rm outside},2}^{2}$ ratio is about 1.3 for PGCM and 1.5 for NLEFT. The smaller enhancement in $\varepsilon_{\rm out}$ is expected, since $R^2_{\rm out}$ also contains contributions from the emission duration and position-time correlation, which dilute the geometric signal. Toward more peripheral collisions the enhancement decreases, and by 20--30\% centrality the \textit{ab initio} results approach the W-S baseline. All three observables clearly separate the \textit{ab initio} configurations from the W-S baseline in central collisions. The difference between NLEFT and PGCM is most significant in the $2R_{{\rm outside},2}^{2}$ ratio, whereas in the $\varepsilon_{\rm side}$ ratio their uncertainty bands overlap in most centrality classes.

For comparison, the $\rm Ne{+}Ne/O{+}O$ ratio of the two-particle elliptic flow coefficient $v_2\{2\}$ in the same simulations reaches 1.14 for PGCM and 1.21 for NLEFT in 0--5\% centrality~\cite{Karmakar:2026inprep}, consistent with previous model comparisons accompanying the measured ratios~\cite{Giacalone:2024luz,ALICE:2025luc,CMS:2025tga}, while it is 0.96--0.98 for ${\textnormal{W-S}}$. The ordering of the configurations is thus the same in both observables, suggesting that the femtoscopic enhancement is driven by the same geometric features of $^{20}$Ne and $^{16}$O that govern the elliptic flow. While $v_2\{2\}$ corresponds to the root-mean-square of the event-by-event elliptic flow, $\varepsilon_{\rm side}$ is extracted relative to the event plane and thus averages the event-by-event fluctuations differently. Although the comparison is therefore not strictly like-for-like, the deviation of the $\varepsilon_{\rm side}$ ratio from unity is still considerably larger than that of the $v_2\{2\}$ ratio.

To test the sensitivity to early-time dynamics, Fig.~\ref{fig:centdepratios} also shows the PGCM configuration without pre-equilibrium free streaming. Removing free streaming changes the absolute source parameters considerably (see Supplemental Material), but the $\rm Ne{+}Ne/O{+}O$ ratios of PGCM and PGCM w/o FS are consistent within uncertainties. The $v_2\{2\}$ ratio is similarly insensitive to the removal of free streaming~\cite{Karmakar:2026inprep}, so both observables retain the geometric signal under this variation, with a considerably larger magnitude in the femtoscopic case.

\begin{figure}
    \centering
    \includegraphics[width=\linewidth]{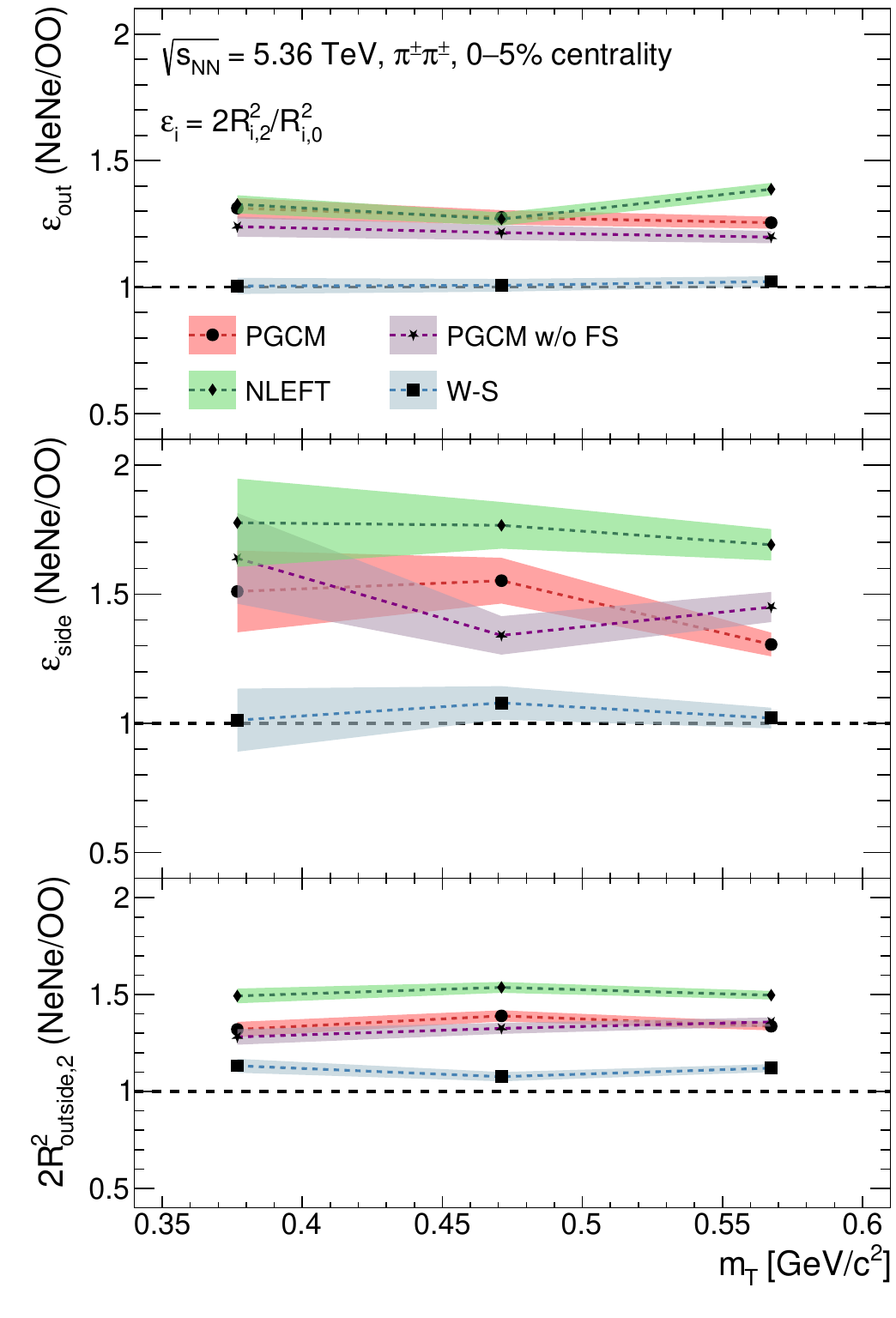}
    \caption{Average transverse mass dependent ratios of the second-order Fourier terms of the azimuthally sensitive femtoscopic radii in Ne+Ne to O+O collisions. The panels display the relative eccentricities $\varepsilon_{\text{out}}$ (top) and $\varepsilon_{\text{side}}$ (middle), alongside the cross-term $2R_{\text{outside},2}^2$ (bottom). On each panel, the initial nucleon configuration setups are shown with different marker shapes and colored bands representing uncertainties propagated from those of the individual quantities.}
    \label{fig:mtdepratios}
\end{figure}

Fig.~\ref{fig:mtdepratios} shows the $m_T$ dependence of the ratios in 0--5\% centrality. The W-S ratios remain close to unity at all $m_T$, and the enhancement for the \textit{ab initio} configurations persists over the studied $m_T$ range. Because radial flow causes pairs at different $m_T$ to probe different homogeneity regions, this indicates that the imprint of nuclear geometry is not confined to a particular part of the freeze-out region. The separation between NLEFT and PGCM is also stable across $m_T$ in the $2R_{{\rm outside},2}^{2}$ ratio, and in this most central class it is visible in the $\varepsilon_{\rm side}$ ratio as well, confirming that the femtoscopic observables can discriminate between the two \textit{ab initio} models.

\paragraph{\label{sec:sum}Summary and outlook.} In summary, we have studied azimuthally sensitive three-dimensional pion femtoscopy in $^{20}\rm Ne{+}^{20}\rm Ne$ and $^{16}\rm O{+}^{16}\rm O$ collisions at $\sqrt{s_{NN}} = 5.36$~TeV, using a hybrid framework of \textit{ab initio} nuclear configurations, viscous hydrodynamics, and hadronic transport. From fully three-dimensional L\'evy-stable fits to the simulated correlation functions, we extracted the azimuthal dependence of the source radii relative to the event plane, and constructed the transverse freeze-out eccentricities $\varepsilon_{\rm side}$ and $\varepsilon_{\rm out}$, as well as the out-side cross-term amplitude $2R^2_{{\rm outside},2}$. The $\rm Ne{+}Ne/O{+}O$ ratios of these quantities are close to unity for a spherical Woods-Saxon baseline, but are strongly enhanced in central collisions for the PGCM and NLEFT configurations resembling $\alpha$-clustering, by about 50--80\% for $\varepsilon_{\rm side}$. This is several times the corresponding enhancement of the $v_2\{2\}$ ratio in the same simulations, while the ordering of the configurations is the same in both observables. The cross-term ratio, in addition, clearly separates the two \textit{ab initio} models. These enhancements persist across the studied $m_T$ range and remain stable when pre-equilibrium free streaming is removed, even though the absolute source radii change considerably. Azimuthally sensitive femtoscopy thus provides a coordinate-space probe of nuclear geometry with enhanced sensitivity, complementary to anisotropic flow.

We emphasize the experimental feasibility of extracting these geometric signatures. Azimuthally differential femtoscopic analyses have been performed at the LHC~\cite{ALICE:2017gxt}, and L\'evy-stable source parameters have been extracted in azimuthally integrated measurements~\cite{CMS:2023xyd}. The recently developed self-consistent treatment of Coulomb effects for three-dimensional L\'evy-stable sources~\cite{Nagy:2026mvg} now provides the mathematical framework to unify these techniques, enabling fully azimuthally sensitive 3D Lévy measurements. The large $\rm O{+}O$ and $\rm Ne{+}Ne$ data samples recently collected at the LHC make such measurements promising. The eccentricity ratios predicted in this work provide a baseline that can be confronted with these forthcoming high-precision measurements.

\paragraph{\label{sec:ack}Acknowledgements.}
The authors are grateful to Máté Csanád and Márton I. Nagy for useful discussions.
DK was supported by the Hungarian NKFIH grants PD-146589 and 2021-4.1.2-NEMZ\_KI-2024-00053, and the Visegrad Fund fellowship No. 62610041. BK, IK and BT acknowledge the support from the Czech Science Foundation, under No 25-16877S. Computational resources were provided by the e-INFRA CZ project (ID:90254), supported by the Ministry of Education, Youth and Sports of the Czech Republic.



\bibliography{ref}

\appendix
\setcounter{figure}{0}
\renewcommand{\thefigure}{S\arabic{figure}}
\section*{Supplemental material}

\begin{figure*}
    \centering
    \includegraphics[width=\linewidth]{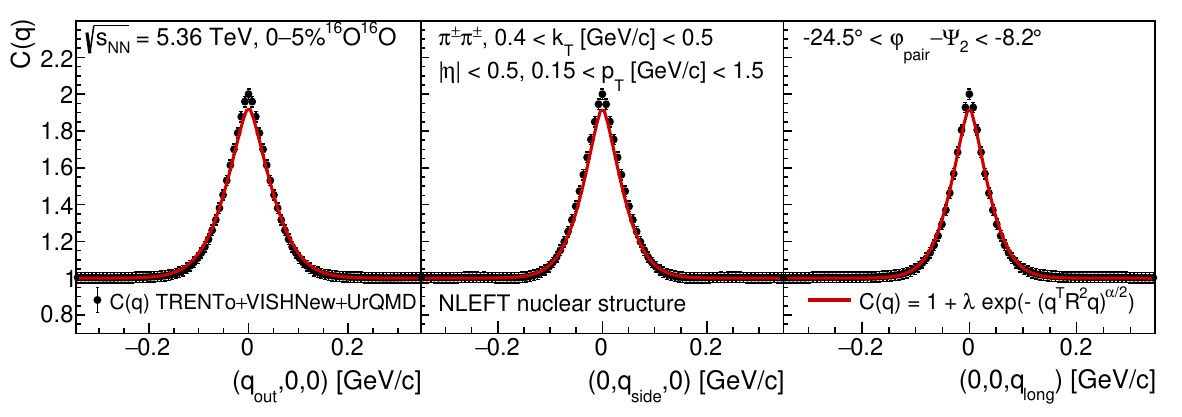}
    \caption{Example fit to a three-dimensional momentum correlation function, illustrated with slices along the three principal axes. The correlation function reconstructed from the simulations and its rescaled statistical uncertainties are shown with black markers and error bars; the fit of Eq.~\eqref{e:Cq3D} is shown with a red line. The collision system and the single-track and pair selection criteria are indicated in the panels.}
    \label{fig:fitexample}
\end{figure*}

In this Supplemental Material, we illustrate the methodological steps leading to the ratios presented in Figs.~\ref{fig:centdepratios} and \ref{fig:mtdepratios} of the main text. Fig.~\ref{fig:fitexample} shows an example of a fully three-dimensional fit of Eq.~\eqref{e:Cq3D} to a correlation function reconstructed from the simulations. The pair-distance distribution $D(\vec\rho)$ is histogrammed on a grid of $211^3$ bins, covering $|\rho_i| < 85$~fm; we verified that the extracted parameters are stable against variations of the grid size and binning. Because $C_2(\vec{q})$ is obtained by a fast Fourier transform of the histogrammed $D(\vec{\rho})$, each bin of $C_2(\vec{q})$ is a linear combination of all coordinate-space bins, and the bin uncertainties are correlated. For statistically independent pairs, the variance of each bin would be $C_2(2\vec{q})/(2N_{\rm pairs})$, i.e., uniform outside the correlation peak and larger by at most a factor of $1+\lambda$ within a region of half the peak width. The overall scale of the uncertainties is, however, not directly determined, since the pairs are not statistically independent: each pion contributes to many pairs, and event-by-event fluctuations of the source within a centrality class add to the counting fluctuations, as do the correlations between oversampled events. We therefore fit with uniform uncertainties, which does not affect the central values for a given weighting, and rescale the parameter uncertainties of each fit by $\sqrt{\chi^2/\mathrm{NDF}}$, following the scale-factor prescription of the Particle Data Group (Sec.~5.2.2 of the Introduction of Ref.~\cite{ParticleDataGroup:2026mpi}). Since bin-to-bin correlations are not explicitly taken into account, the resulting uncertainties should be regarded as estimates. A more rigorous determination, accounting for all of the above correlations, would be provided by a jackknife or bootstrap resampling over hydrodynamic events, each taken together with its oversampled events; this is left for future work owing to its computational cost for the large number of three-dimensional fits performed here.

\begin{figure*}
    \centering
    \includegraphics[width=0.49\linewidth]{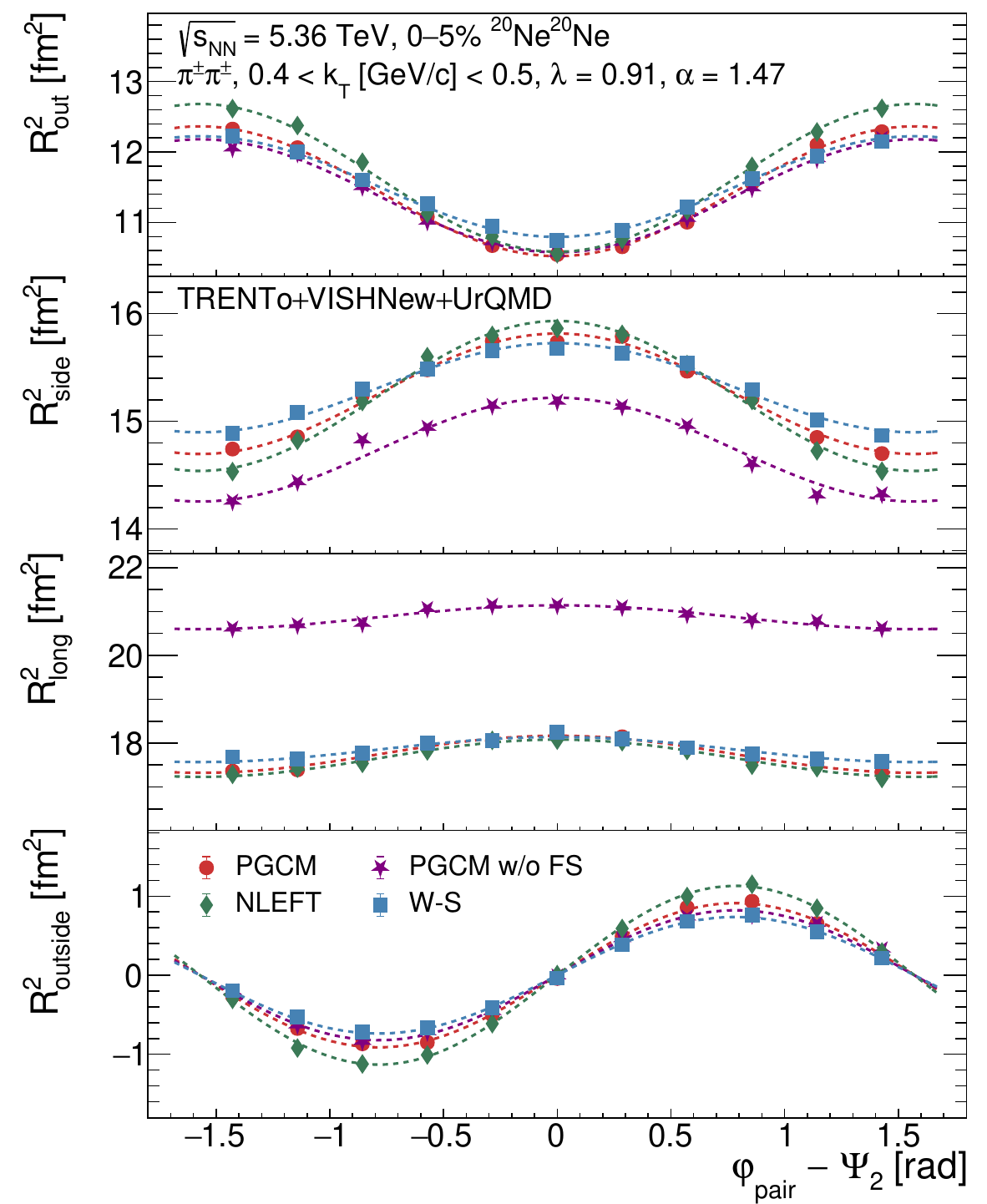}
    \includegraphics[width=0.49\linewidth]{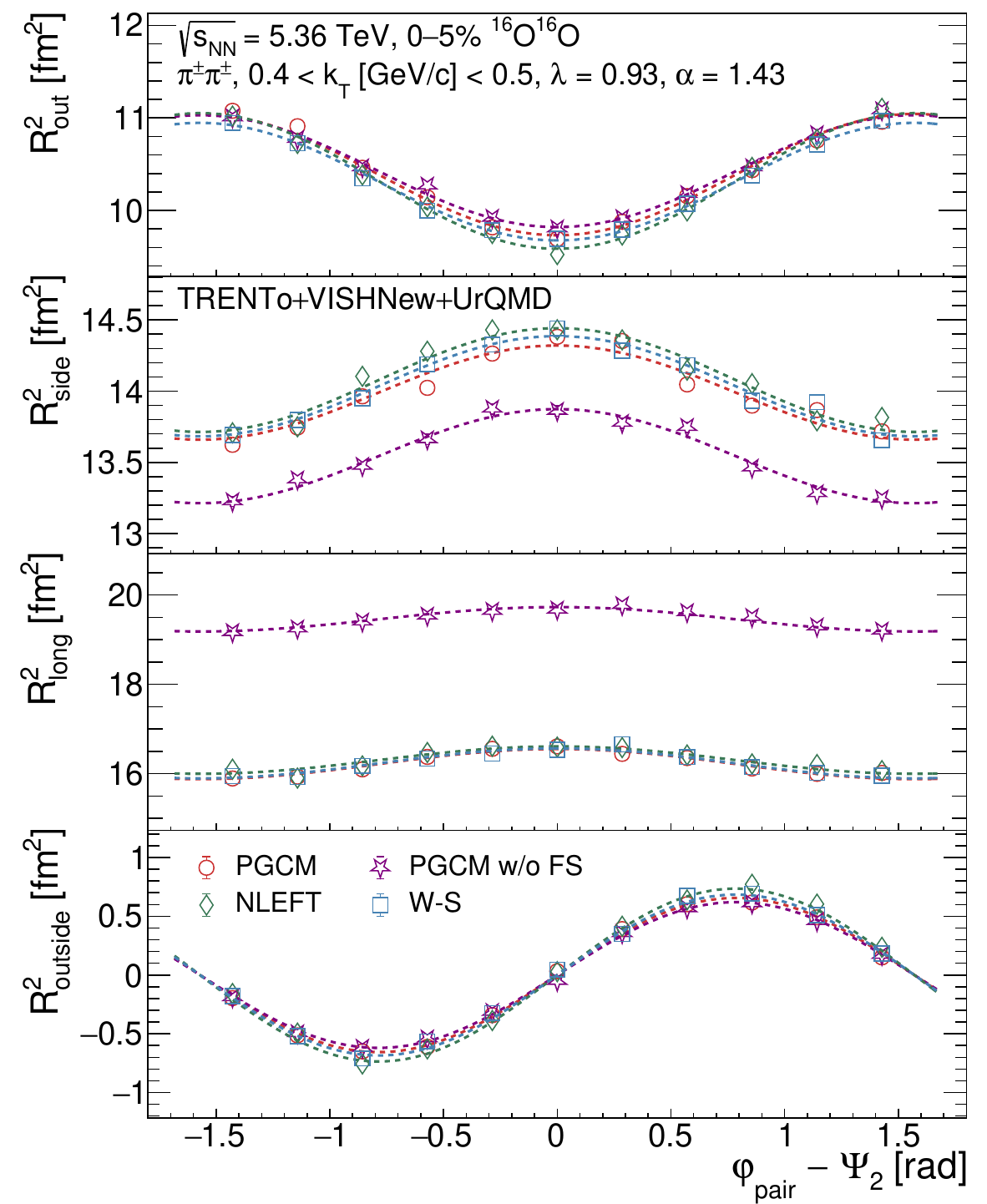}
    \caption{Examples of Fourier series fits to the azimuthal dependence of the extracted squared radii in 0--5\% $\rm Ne{+}Ne$ (left) and $\rm O{+}O$ (right) collisions. From top to bottom, the panels show $R_{\rm out}^2$, $R_{\rm side}^2$, $R_{\rm long}^2$, and $R_{\rm outside}^2$. The four initial-state configurations are shown with different colors and markers. The error bars represent the statistical uncertainties from the correlation function fits, and the dashed lines show the fits of Eq.~\eqref{e:oscillation}.}
    \label{fig:oscillation}
\end{figure*}

These fits are performed in all intervals of centrality, $k_T$, and $\varphi_{\rm pair}-\Psi_2$. Fig.~\ref{fig:oscillation} shows examples of the Fourier series fits of Eq.~\eqref{e:oscillation} to the azimuthal dependence of the extracted radii. In practice, Eq.~\eqref{e:oscillation} is fitted in the equivalent form $$R_i^2(\varphi_{\rm pair}-\Psi_2) = R_{i,0}^2\left[1+\varepsilon_i\cos(2(\varphi_{\rm pair}-\Psi_2))\right]$$ for $i\in\{{\rm out},{\rm side},{\rm long}\}$, so that $\varepsilon_i$ and its uncertainty are obtained directly from the fit. Since $\lambda$ and the L\'evy exponent $\alpha$ show no significant dependence on $\varphi_{\rm pair}-\Psi_2$, we use a two-step procedure: we first fit $C_2(\vec{q})$ with $\lambda$ and $\alpha$ as free parameters in all azimuthal bins and compute their azimuthal averages, and then repeat the fits with $\lambda$ and $\alpha$ fixed to these averages. This reduces the bin-to-bin fluctuations of the radii induced by parameter correlations and improves the precision of the extracted oscillations. The $\lambda$ and $\alpha$ values indicated in Fig.~\ref{fig:oscillation} are these azimuthal averages. This method would be viable for experiments as well, where fixing these parameters to values determined from an azimuthally integrated analysis would also decrease the radii uncertainties.

\begin{figure*}
    \centering
    \includegraphics[width=0.49\linewidth]{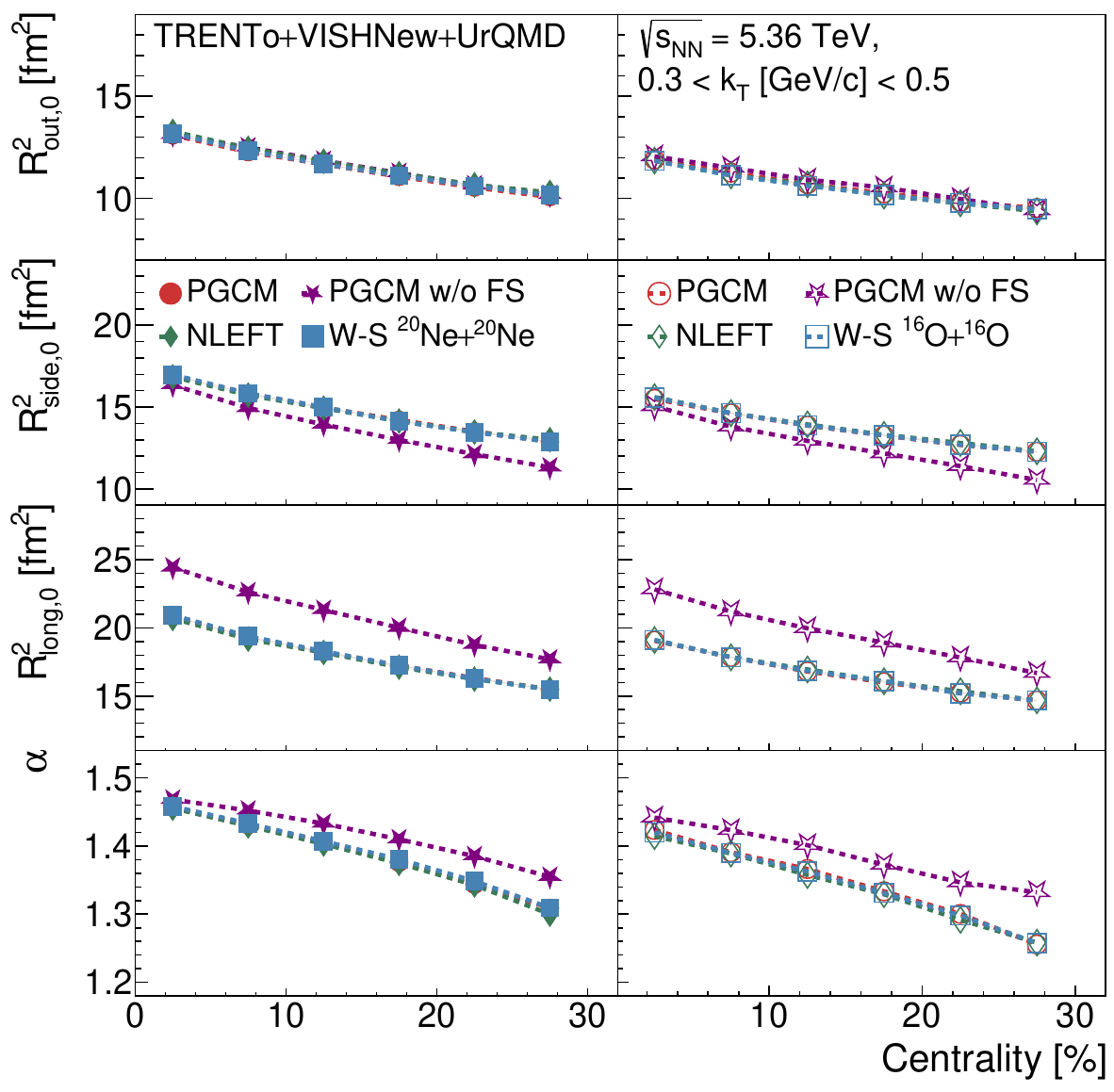}
    \includegraphics[width=0.49\linewidth]{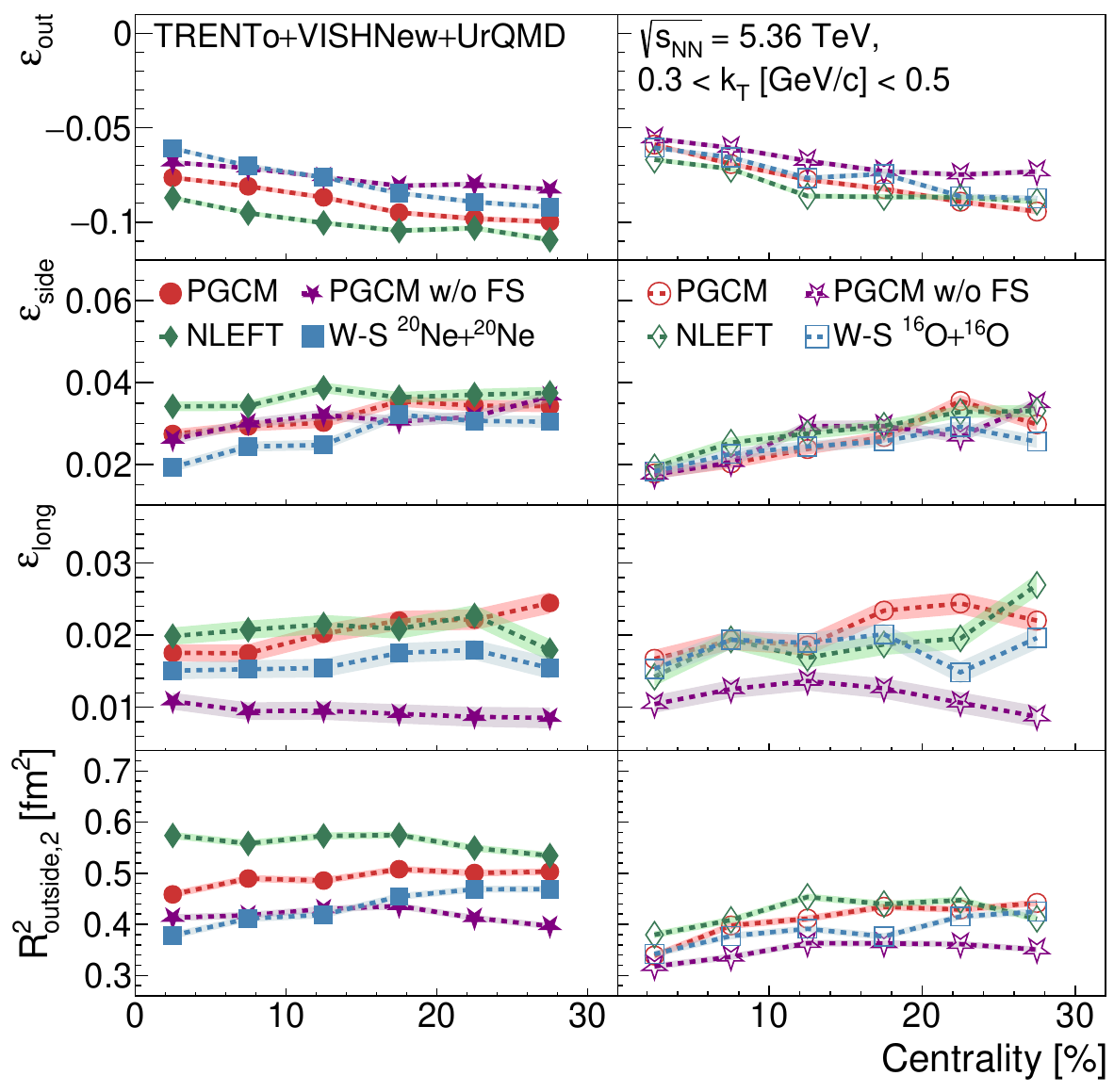}
    \caption{Centrality dependence of the source parameters for $0.3 < k_T < 0.5$~GeV/$c$. The left two columns show the zeroth-order Fourier terms $R^2_{{\rm out},0}$, $R^2_{{\rm side},0}$, $R^2_{{\rm long},0}$, and the L\'evy exponent $\alpha$; the right two columns show the second-order terms $\varepsilon_{\rm out}$, $\varepsilon_{\rm side}$, $\varepsilon_{\rm long}$, and $R^2_{{\rm outside},2}$. Filled (empty) markers denote $\rm Ne{+}Ne$ ($\rm O{+}O$) collisions for the four initial-state configurations. The bands represent the statistical uncertainties from the oscillation fits.}
    \label{fig:centdep_params}
\end{figure*}

\begin{figure*}
    \centering
    \includegraphics[width=0.49\linewidth]{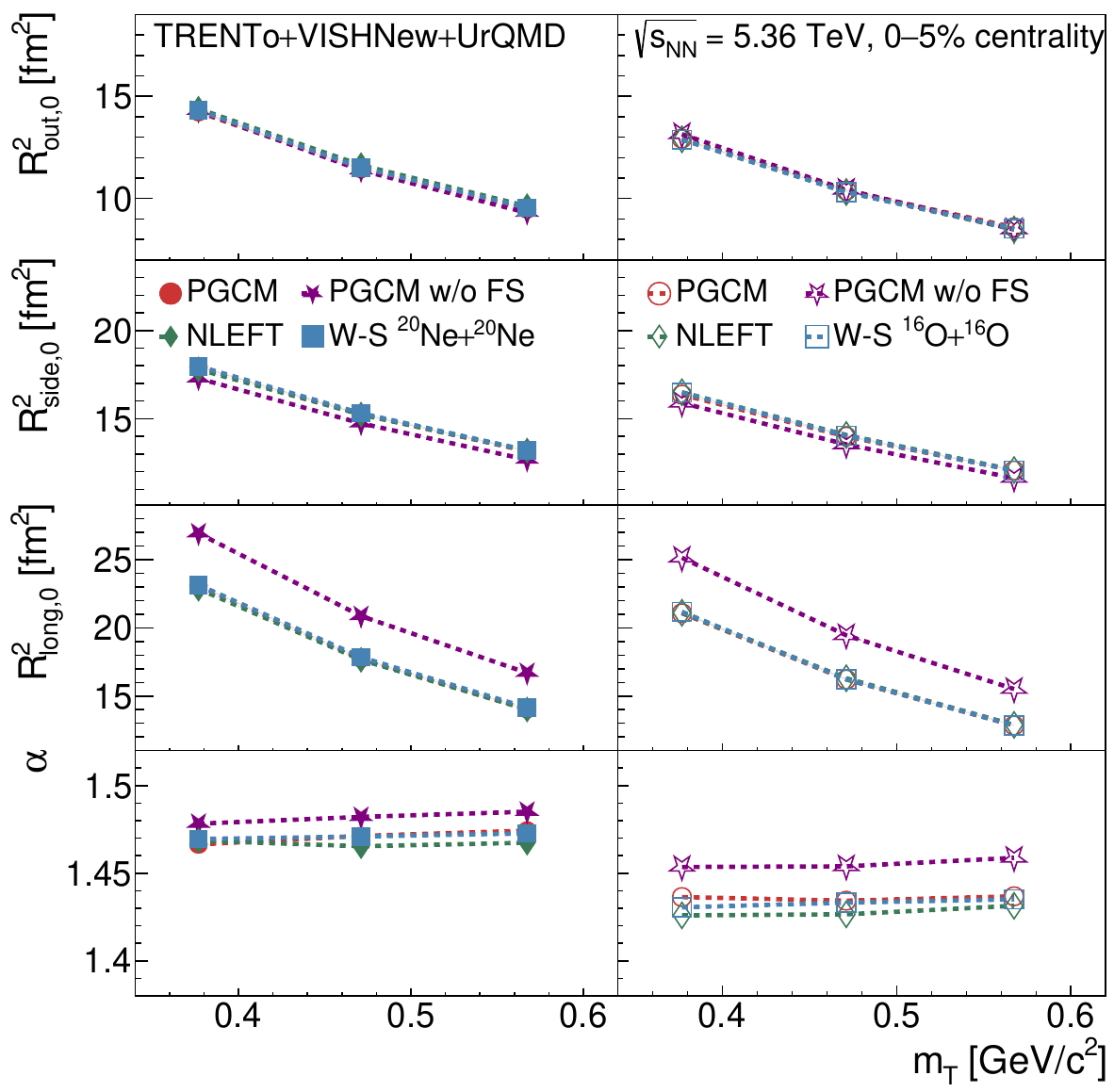}
    \includegraphics[width=0.49\linewidth]{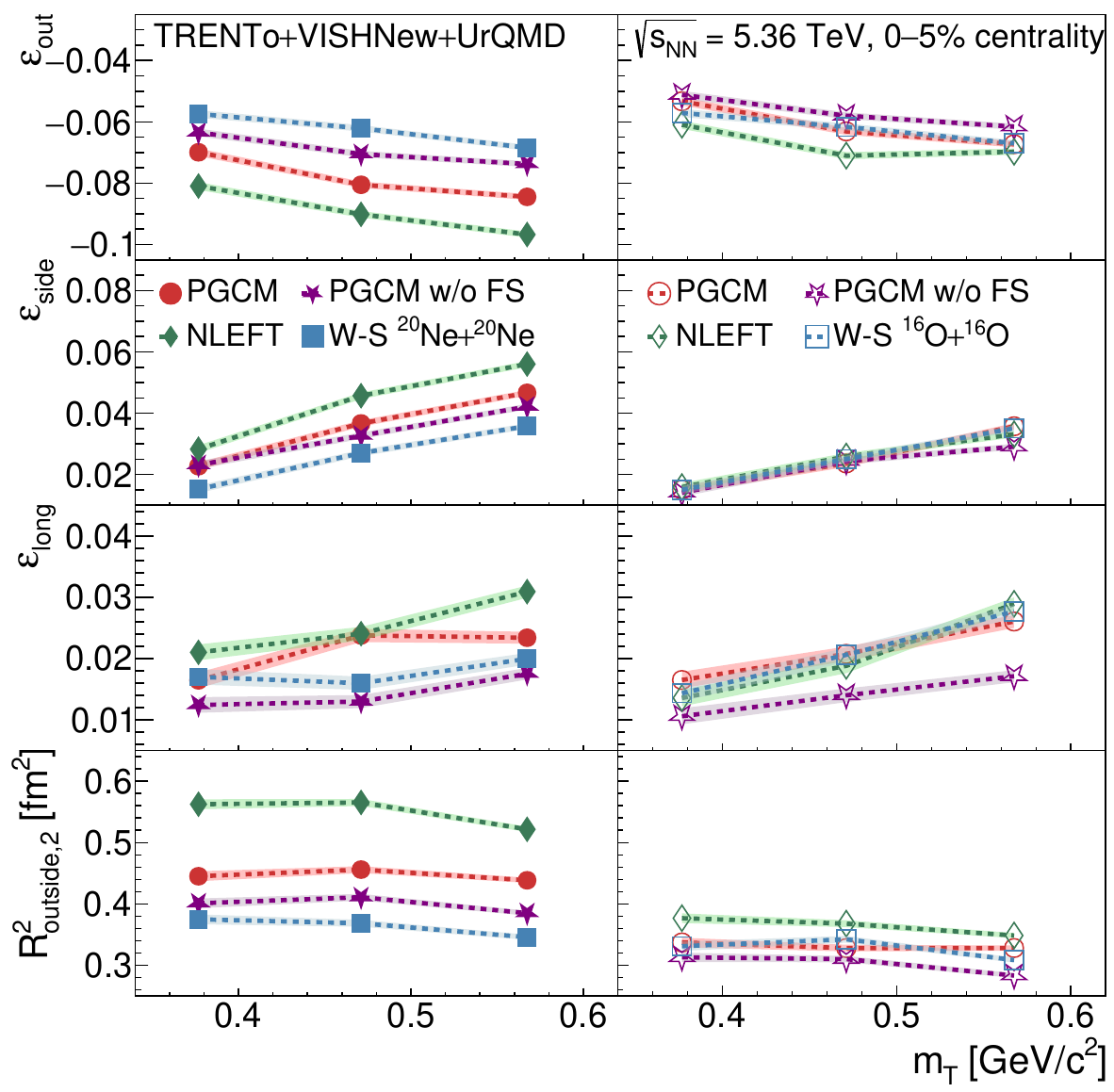}
    \caption{Transverse-mass dependence of the source parameters in 0--5\% centrality. The left two columns show the zeroth-order Fourier terms $R^2_{{\rm out},0}$, $R^2_{{\rm side},0}$, $R^2_{{\rm long},0}$, and the L\'evy exponent $\alpha$; the right two columns show the second-order terms $\varepsilon_{\rm out}$, $\varepsilon_{\rm side}$, $\varepsilon_{\rm long}$, and $R^2_{{\rm outside},2}$. Filled (empty) markers denote $\rm Ne{+}Ne$ ($\rm O{+}O$) collisions for the four initial-state configurations. The bands represent the statistical uncertainties from the oscillation fits.}
    \label{fig:mtdep_params}
\end{figure*}

The resulting zeroth- and second-order Fourier terms are shown as a function of centrality in Fig.~\ref{fig:centdep_params} and of $m_T$ in Fig.~\ref{fig:mtdep_params}. The correlation strength $\lambda$ is not shown: since the UrQMD cascade is terminated at a finite time of 8000~fm/$c$, the simulated source does not include pions from weak decays of long-lived particles (such as $K^0_S$ or $\Lambda$), so $\lambda$ does not correspond to the experimental intercept parameter and serves only as an effective normalization in the fit, similarly to Refs.~\cite{Molnar:2026vgi,Porfy:2025sev}. Figs.~\ref{fig:centdep_params} and~\ref{fig:mtdep_params} also show that the absolute source parameters, in particular $R^2_{{\rm long},0}$, $\alpha$, and $\varepsilon_{\rm long}$, depend considerably on the inclusion of pre-equilibrium free streaming. This sensitivity is largely canceled in the ratios, but makes the absolute source parameters a promising input for future Bayesian analyses of the initial-state and pre-equilibrium dynamics in light-ion collisions.


\end{document}